\documentclass[final,1p,times]{elsarticle} 
\usepackage{graphicx} 
\usepackage{amssymb} 
\usepackage{amsthm} 
\usepackage{lineno} 

\usepackage{color}
\def\mop#1{\mathop{\rm #1}\nolimits}
\def\tr{\mop{tr}}
\def\lsim{\mathrel{\mathstrut\smash{\ooalign{\raise2.5pt\hbox{$<$}\cr\lower2.5pt\hbox{$\sim$}}}}}
\def\gsim{\mathrel{\mathstrut\smash{\ooalign{\raise2.5pt\hbox{$>$}\cr\lower2.5pt\hbox{$\sim$}}}}}

\journal{Nuclear Physics A} 
\begin{document} 

\begin{frontmatter} 


\title{Using string theory to study the quark-gluon plasma: progress and perils}

\author{Steven S. Gubser}

\address{Joseph Henry Laboratories, Princeton University, Princeton, NJ  08544}

\begin{abstract} 
I outline some of the progress over the past few years in applying ideas from string theory to study the quark-gluon plasma, including the computation of the drag force on heavy quarks and estimates of total multiplicity from black hole formation.  I also indicate some of the main perils of the string theory approach.

\end{abstract} 

\end{frontmatter} 

\linenumbers 

\section{Introduction}\label{INTRODUCTION}

The gauge-string duality \cite{Maldacena:1997re,Gubser:1998bc,Witten:1998qj} provides a powerful computational framework for studying gauge theories at strong coupling.  Recent years have seen a focused effort to use this duality to understand the quark-gluon plasma, despite the obvious differences between the best understood string theory constructions and quantum chromodynamics (QCD).

In the gauge-string duality, a thermal state analogous to the quark-gluon plasma is represented as a black hole.  Five-dimensional calculations based on the properties of the black hole horizon are translated into statements about strongly coupled thermal gauge theories, using the gauge-string duality.  These statements can then be compared with measured properties of the quark-gluon plasma.

In this contribution I will highlight a selection of encouraging results from the string theory approach, and at the same time point out the weak points of the calculations behind these results.

The organization of the rest of this contribution is as follows.  In section~\ref{GAUGESTRING} I will introduce the gauge-string duality.  In section~\ref{HARDPROBES} I will discuss predictions related to hard probes.  And in section~\ref{BULK} I will discuss predictions for bulk physics.  Broader reviews of the gauge-string duality and its applications to QCD include \cite{MAGOO,Klebanov:2000me,Gubser:2009md,Gubser:2009sn}.  For a recent and useful review of string theory constructions more specifically tailored to fit QCD than SYM, see \cite{KiritsisTalk}.

\section{The gauge-string duality}\label{GAUGESTRING}

\subsection{${\cal N}=4$ super-Yang-Mills is dual to $AdS_5 \times S^5$ \cite{Maldacena:1997re,Gubser:1998bc,Witten:1998qj}}

\begin{figure}[ht]
  \centering\includegraphics[width=2.5in]{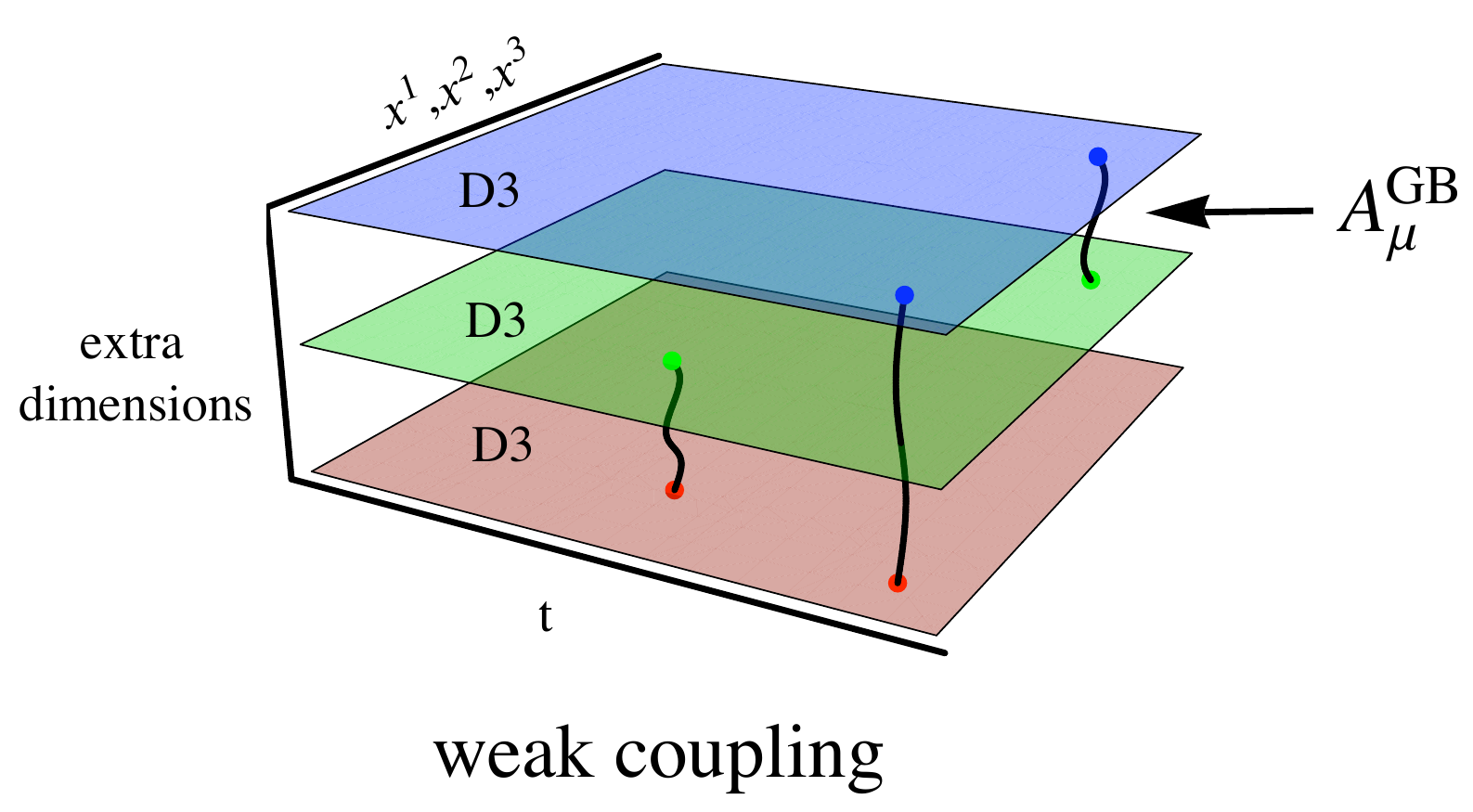}\qquad\qquad\includegraphics[width=2in]{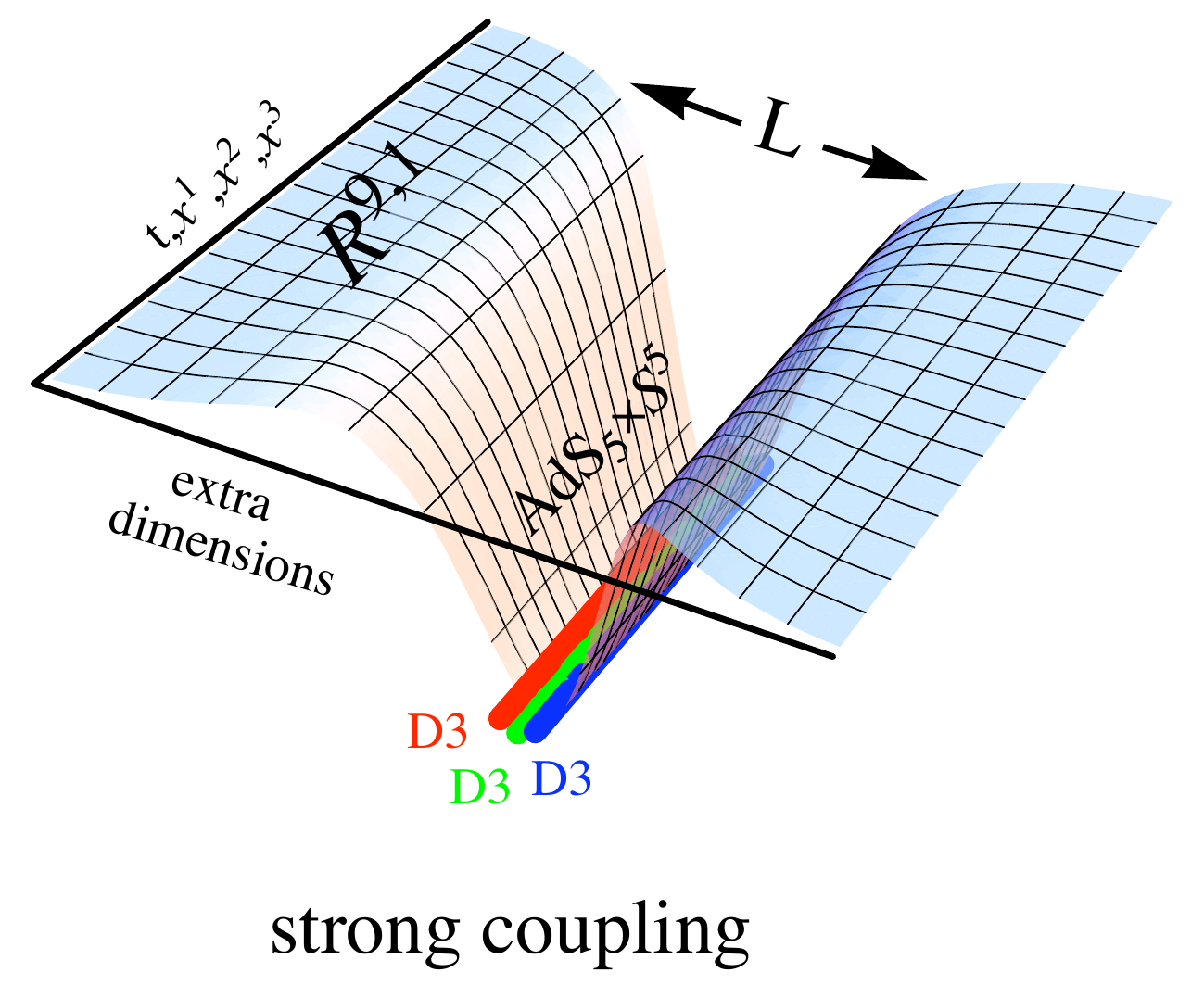}
  \caption[]{(Color online.)  D3-branes at weak coupling and at strong coupling.  At weak coupling, strings stretched between the branes behave as nearly free gluons in a 3+1-dimensional gauge theory, ${\cal N}=4$ SYM.  At strong coupling, the curved geometry $AdS_5 \times S^5$ captures the gauge theory dynamics.}\label{WeakStrong}
 \end{figure}

This duality, the simplest example of AdS/CFT, comes from two apparently different ways of describing D3-branes, which are locations in ten-dimensional spacetime where strings can end (see figure~\ref{WeakStrong}).    
Low-energy excitations of D3-branes are governed by ${\cal N}=4$ SYM, whose lagrangian takes the schematic form
 \begin{equation}\label{LSYM}
  {\cal L}_{\rm SYM} -{1 \over 2 g_{YM}^2} \tr F^2 + \hbox{(superpartners)} \,.
 \end{equation}
$\beta(g_{YM}) \equiv 0$ for this theory, which makes comparison with QCD {\bf perilous}.

At weak coupling, the open strings ending on the D3-branes behave as gluons, and superpartners of gluons, in SYM.  At strong coupling, the simplest description of D3-branes is in terms of near-horizon geometry, $AdS_5 \times S^5$.  The five-sphere, $S^5$, plays essentially no role in the discussion to follow.  The metric of $AdS_5$ is
 \begin{equation}\label{AdSMet}
  ds^2 = {L^2 \over z^2} (-dt^2 + d\vec{x}^2 + dz^2) \qquad\qquad\hbox{($z>0$)} \,,
 \end{equation}
and the characteristic length scale $L$ is related to the 't~Hooft coupling by
 \begin{equation}\label{Lrelation}
  \lambda \equiv g_{YM}^2 N = {L^2 \over \alpha'} \,,
 \end{equation}
where $\alpha'$ is the inverse string tension.  Strong coupling means $\lambda \gg 1$, which is equivalent to $L^2 \gg \alpha'$.  This is the statement that strings are typically much smaller than the radius of curvature of $AdS_5$, making geometrical notions like the metric reliable.

\subsection{The equation of state has $\displaystyle{\epsilon \over \epsilon_{\rm free}} = {3 \over 4} + {1.69 \over \lambda^{3/2}} + \ldots$ \cite{Gubser:1996de,Gubser:1998nz,Blaizot:2006tk}}

One can introduce finite temperature by replacing $AdS_5$ by the $AdS_5$-Schwarzschild metric:
 \begin{equation}\label{AdSSch}
  ds^2 = {L^2 \over z^2} \left( -h dt^2 + d\vec{x}^2 + {dz^2 \over h} \right)
   \qquad\qquad h = 1-{z^4 \over z_H^4} \,.
 \end{equation}
The horizon at $z=z_H$ has Hawking temperature $T = 1/(\pi z_H)$, and its Bekenstein-Hawking entropy is related to its area by $S = A / 4 G_N$.

SYM has considerably more degrees of freedom than QCD: free field counting gives
 \begin{equation}\label{DoF}
  \epsilon_{\rm free}^{SYM} \approx 39 T^4 \qquad\qquad
   \epsilon_{\rm free}^{QCD} \approx 16 T^4 \quad\hbox{(3 massless flavors)} \,.
 \end{equation}
It is {\bf perilous} to directly compare theories with $\epsilon/T^4$ so different.  On the other hand, it is intriguing that lattice results for $\epsilon/\epsilon_{\rm free}$ in QCD are fairly close to the SYM values.

\subsection{An infinitely massive fundamental quark is dual to a hanging string \cite{Maldacena:1998im,Rey:1998ik}}\label{qqbar}

\begin{figure}
  \centering\includegraphics[width=1.5in,angle=270]{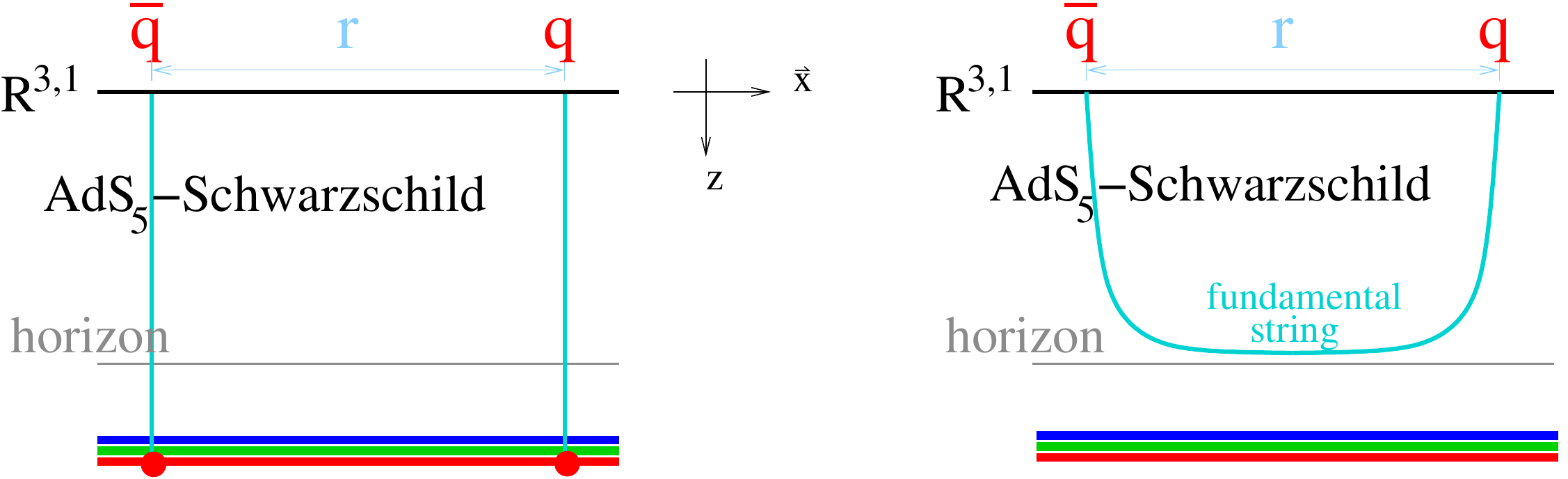}
  \caption[]{(Color online.)  Left: a quark and anti-quark are described on the string theory side by strings hanging from the boundary of $AdS_5$ down to the D3-branes.  Right: A lower-energy configuration is for the two strings to join into one U-shaped string.}\label{QuarkAttraction}
 \end{figure}

The relation between quark-anti-quark pairs and hanging strings in $AdS_5$ is illustrated in figure~\ref{QuarkAttraction}.
The quarks are infinitely massive because there is actually an infinite length of string ``close'' to the boundary.

By calculating on the string theory side the energy gained by passing from the disjoint strings to the U-shaped string, one finds a Coulombic force, screened in the infrared when the temperature is finite.  Equating this force to the quark-anti-quark calculated from the lattice at a separation $r \approx 0.25\,{\rm fm}$ and an energy density corresponding in QCD to $T \approx 240\,{\rm MeV}$ leads to
 \begin{equation}\label{ChooseLambda}
  \lambda_{SYM} = 5.5^{\;+2.5}_{\;-2}
 \end{equation}
in SYM \cite{Gubser:2006qh}.  This is surprising because then $\alpha_{SYM} \approx 0.15$.  The match between lattice calculations and SYM is conspicuously imperfect because SYM doesn't confine.  The leading order string theory curve isn't even fully understood for $r \gsim 0.25\,{\rm fm}$; but see \cite{Bak:2007fk}.  These points illustrate some the {\bf perils} of comparing SYM and lattice QCD; however, I will stick with (\ref{ChooseLambda}) as a physically motivated range of couplings, and also continue to compare SYM and QCD at fixed energy density rather than fixed temperature, as a way of correcting for the larger number of degrees of freedom in SYM.

\section{Hard probes}\label{HARDPROBES}

\subsection{Heavy quarks lose momentum according to $\displaystyle{{dp \over dt} = -{p \over \tau_Q} + \hbox{stochastic}}$ \cite{Herzog:2006gh,Casalderrey-Solana:2006rq,Gubser:2006bz}}

\begin{figure}[ht]
  \centering\includegraphics[width=3.5in]{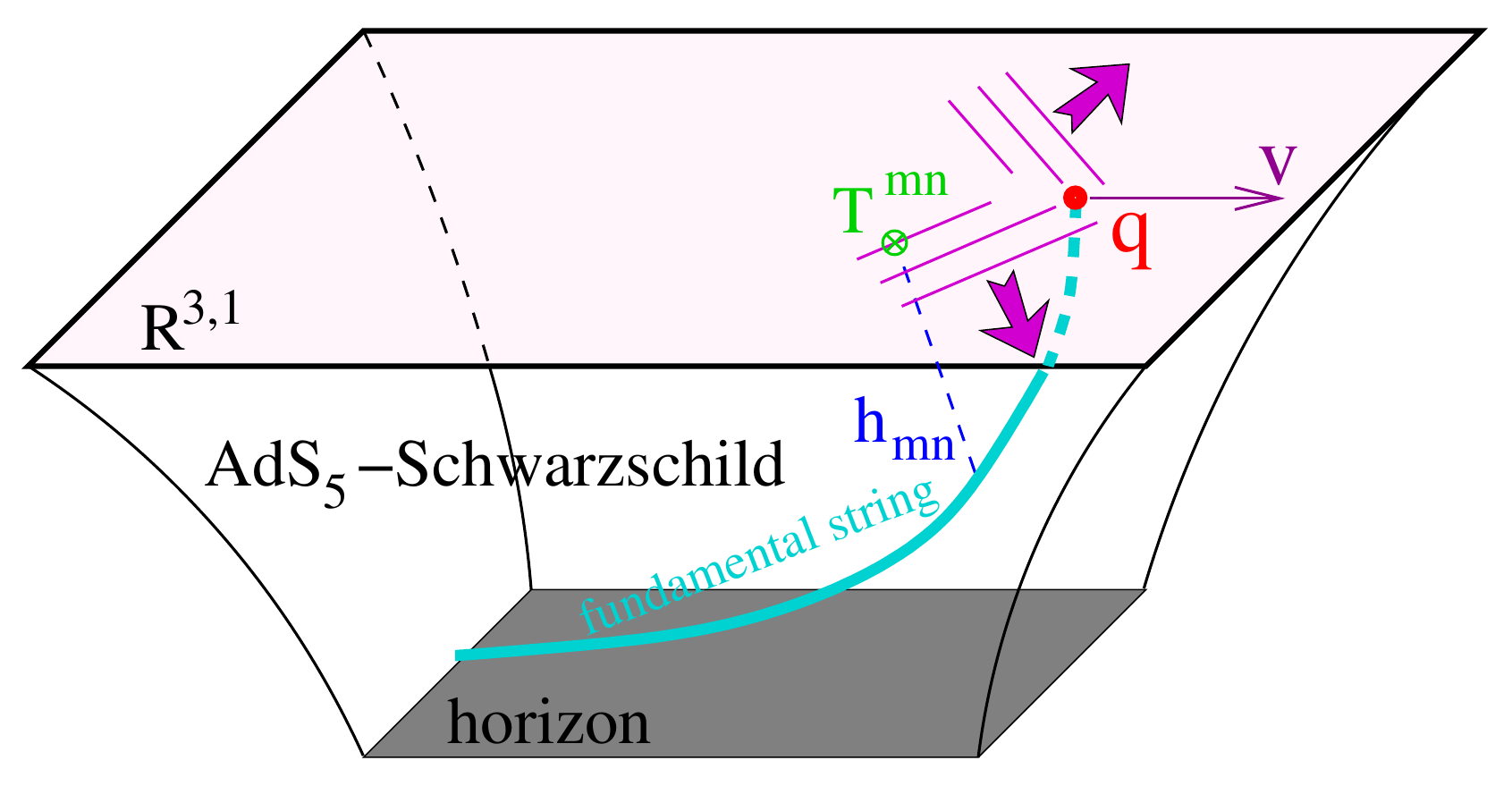}
  \caption[]{(Color online.)  A quark drags a trailing string behind it.  The string encodes energy loss in the dual gauge theory.}\label{Wake}
 \end{figure}

The physical picture of quark energy loss in string theory is sketched in figure~\ref{Wake}.  The trailing string describes the response of the color-electric fields produced by the quark to the thermal medium.
From the shape of the string, determined through classical equations of motion in string theory, one can deduce that
 \begin{equation}\label{dpdt}
  {dp \over dt} = -{\pi\sqrt\lambda \over 2} T_{SYM}^2 
    {v \over \sqrt{1-v^2}} = -{p \over \tau_Q}\qquad\qquad
   \tau_Q = {2m_Q \over \pi T_{SYM}^2 \sqrt\lambda}
 \end{equation}
 \begin{equation}
  \tau_{\rm charm} \approx 2\,{\rm fm} \qquad
  \tau_{\rm bottom} \approx 6\,{\rm fm} \qquad\qquad
    \hbox{if $T_{QCD} = 250\,{\rm MeV}$} \,.
 \end{equation}
A recent study shows that these equilibration times are at least roughly consistent with $R_{AA}$ of non-photonic electrons \cite{Akamatsu:2008ge}.

The stochastic forces on the quark are reflected on the string theory side by oscillations of the trailing string \cite{Casalderrey-Solana:2006rq,Gubser:2006nz,Casalderrey-Solana:2007qw,Giecold:2009cg}.

\subsection{There are two competing efforts to characterize gluon energy loss \cite{Liu:2006ug,Liu:2006he,Gubser:2008as,Hatta:2008tx,Chesler:2008wd,Chesler:2008uy}}

The first method, originally proposed in \cite{Liu:2006ug}, is based on using a Wilson loop describing two quarks separated by a length $L$ and propagating along a null separation $L^-$, measured in the rest frame of the plasma.  In this approach one finds
 \begin{equation}\label{qhatLRW}
  \hat{q}_{\rm LRW} = {\pi^{3/2} \Gamma(3/4) \over \Gamma(5/4)}
    \sqrt\lambda \, T^3 \approx 3.6 \, {{\rm GeV}^2 \over
      {\rm fm}} \left( {T_{SYM} \over 280\,{\rm MeV}} 
       \right)^3 \,.
 \end{equation}
In the last approximate equality, I used $\lambda = 6\pi$, as preferred by the authors of \cite{Liu:2006ug}.  In comparing with QCD, a reduction of $\hat{q}$ by about a factor of $2$ is probably in order to account for fewer degrees of freedom in QCD.  Criticisms of the choice of worldsheet \cite{Argyres:2006yz} suggest that the approach of \cite{Liu:2006ug} is not without its {\bf perils}.

The second method is based on a representation of an off-shell gluon as a string falling into the horizon, first proposed in \cite{Gubser:2008as}.  This method leads to a stopping distance
 \begin{equation}\label{xstop}
  x_{\rm stop} \lsim {1 \over \pi T_{SYM}} \left( {1 \over \sqrt\lambda}
    {E \over T_{SYM}} \right)^{1/3} \,.
 \end{equation}
It is {\bf perilous} to compare (\ref{xstop}) (or better estimates of $x_{\rm stop}$ based on worldsheet and spacetime geodesics) with more standard approaches based on $\hat{q}$, simply because the picture of a gluon advocated in \cite{Gubser:2008as} is significantly different from the perturbative picture.  Nevertheless, one can make a rough translation of $x_{\rm stop}$ into a value for the jet-quenching parameter:
 \begin{equation}\label{qhatRough}
  \hat{q}_{\rm rough} \equiv {4E \over 3\alpha_s x_{\rm stop}^2} 
    \approx 21\,{{\rm GeV}^2 \over {\rm fm}} \,,
 \end{equation}
where in the approximate equality we set $\alpha_s = 1/2$ for QCD, but use $\lambda_{SYM} = 5.5$ for SYM, as well as the usual scheme of matching the energy density of QCD and SYM, with $T_{QCD} = 280\,{\rm MeV}$.  We also assumed that the energy of gluons is between $5$ to $25\,{\rm GeV}$.

Among the {\bf perils} of this discussion are the crudeness of (\ref{qhatRough}), the sensitivity to the temperature, and the obvious tension between the proposals of \cite{Liu:2006ug} and \cite{Gubser:2008as}.

\section{Bulk physics}\label{BULK}

\subsection{In a phenomenological construction, $\zeta/s \,
\textstyle{\lower8pt\hbox{$<$} \atop \raise8pt\hbox{$\sim$}}\, 
1/4\pi$ \cite{Gubser:2008ny,Gubser:2008yx,Gubser:2008sz}}

\begin{figure}[ht]
  \centering\includegraphics[width=2.5in]{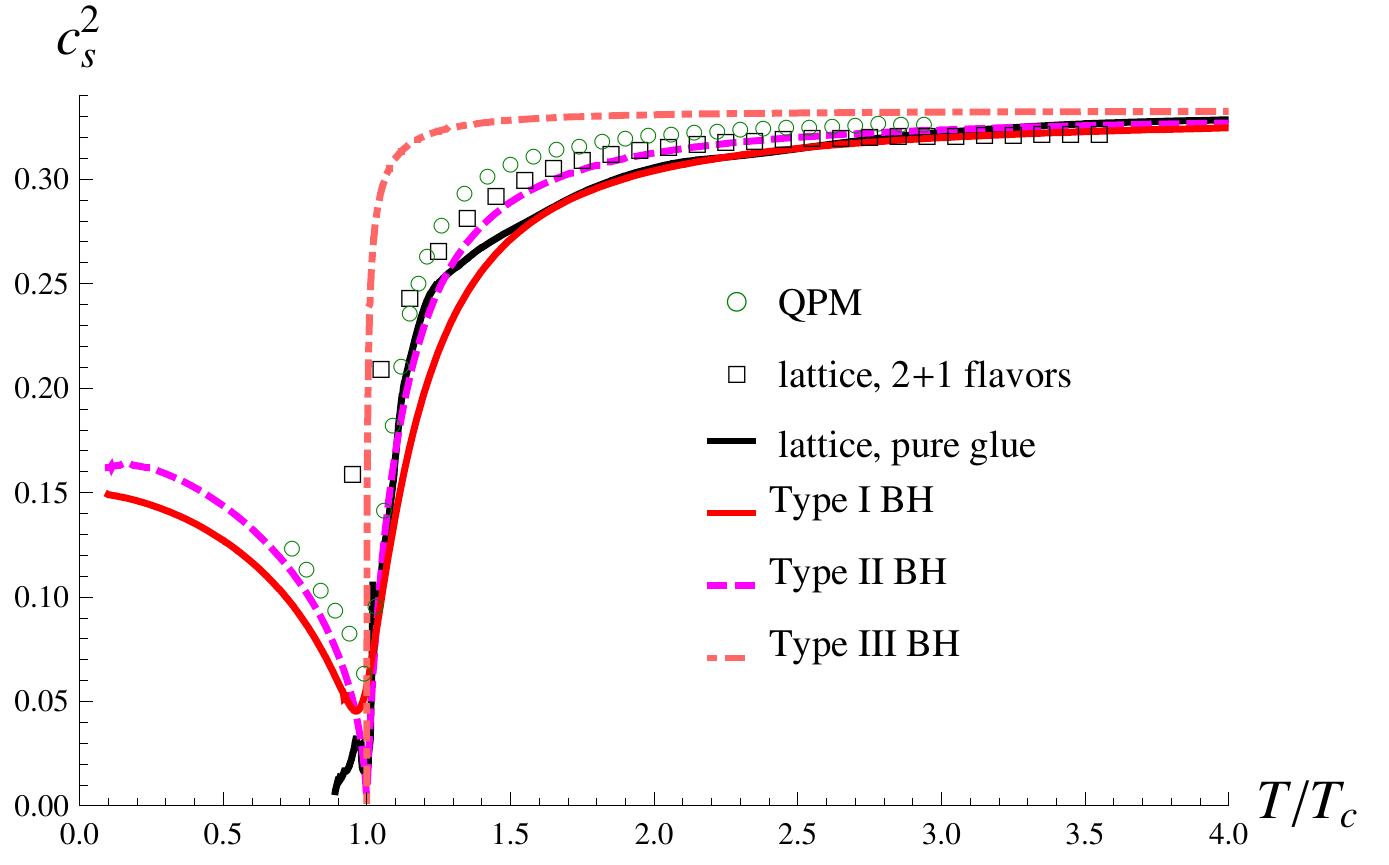}\qquad
   \includegraphics[width=2.5in]{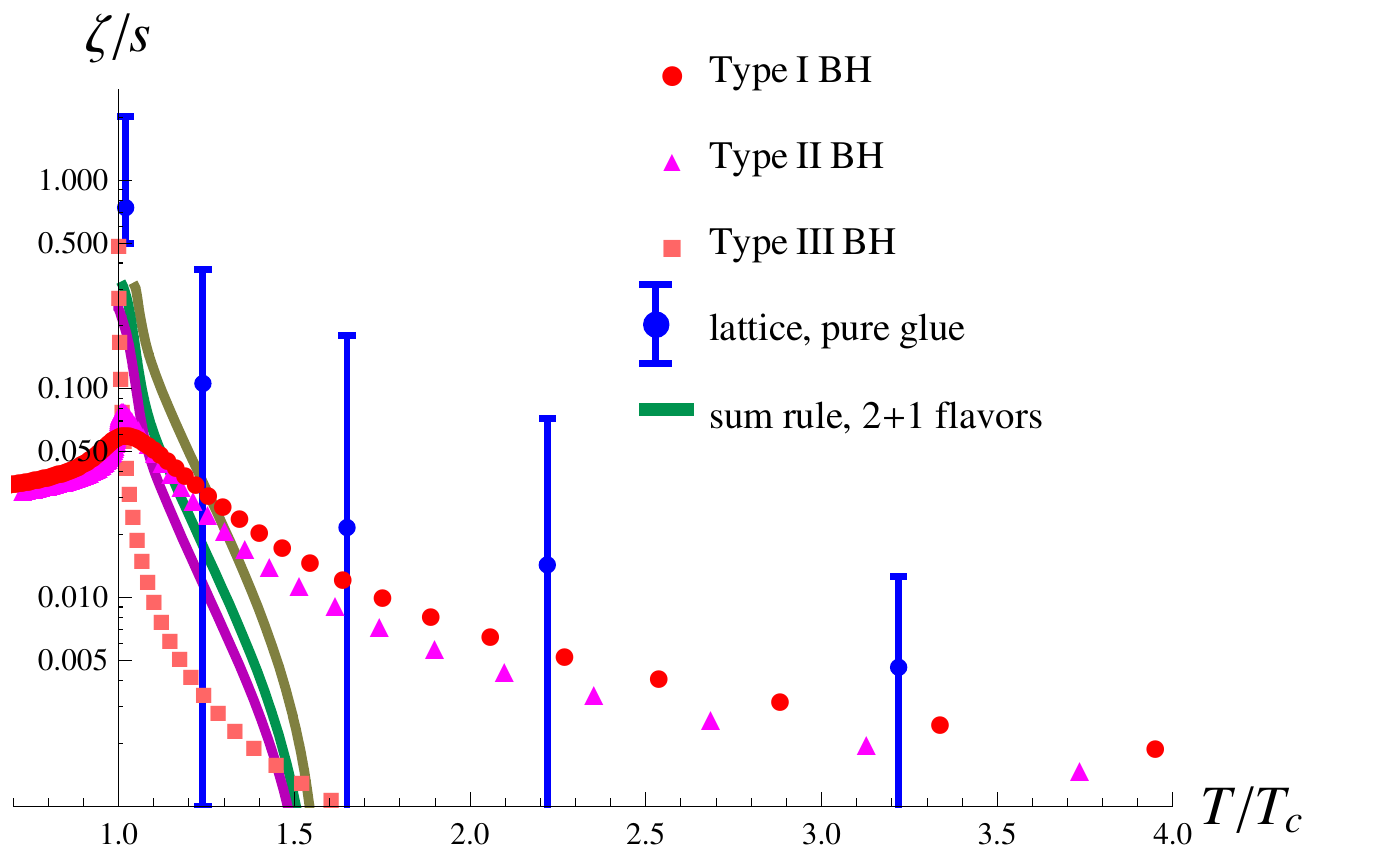}
  \caption[]{(Color online.)  Left: The speed of sound as a function of temperature for several constructions based on (\ref{S5D}).  Right: The corresponding bulk viscosity.  The lattice points, from \cite{Meyer:2007dy}, are for pure glue.  Both plots are from \cite{Gubser:2008yx}.}\label{BulkSummary}
 \end{figure}

In \cite{Gubser:2008ny} it was shown that starting from the gravitational action
 \begin{equation}\label{S5D}
  S_{\rm 5-dimensional} = \int d^5 x \, \sqrt{g} \left[ R - 
   {1 \over 2} (\partial\phi)^2 - V(\phi) \right] \,,
 \end{equation}
one can adjust the scalar potential $V(\phi)$ to mimic the equation of state of QCD.  Having done this, one can calculate the bulk viscosity from the probability for an appropriate superposition of gravitons and scalars to be absorbed by the black hole.  Results of such computations, shown in figure~\ref{BulkSummary}, indicate that the peak in $\zeta/s$ is present, but not strong, for a realistic equation of state.

A {\bf peril} in using the action (\ref{S5D}) is that it doesn't come from a first principles calculation.

\subsection{Multiplicity estimates from colliding shocks have the right magnitude but strange scaling \cite{Gubser:2008pc,Lin:2009pn,Gubser:2009sx}}

\begin{figure}
  \centering\includegraphics[width=2.5in]{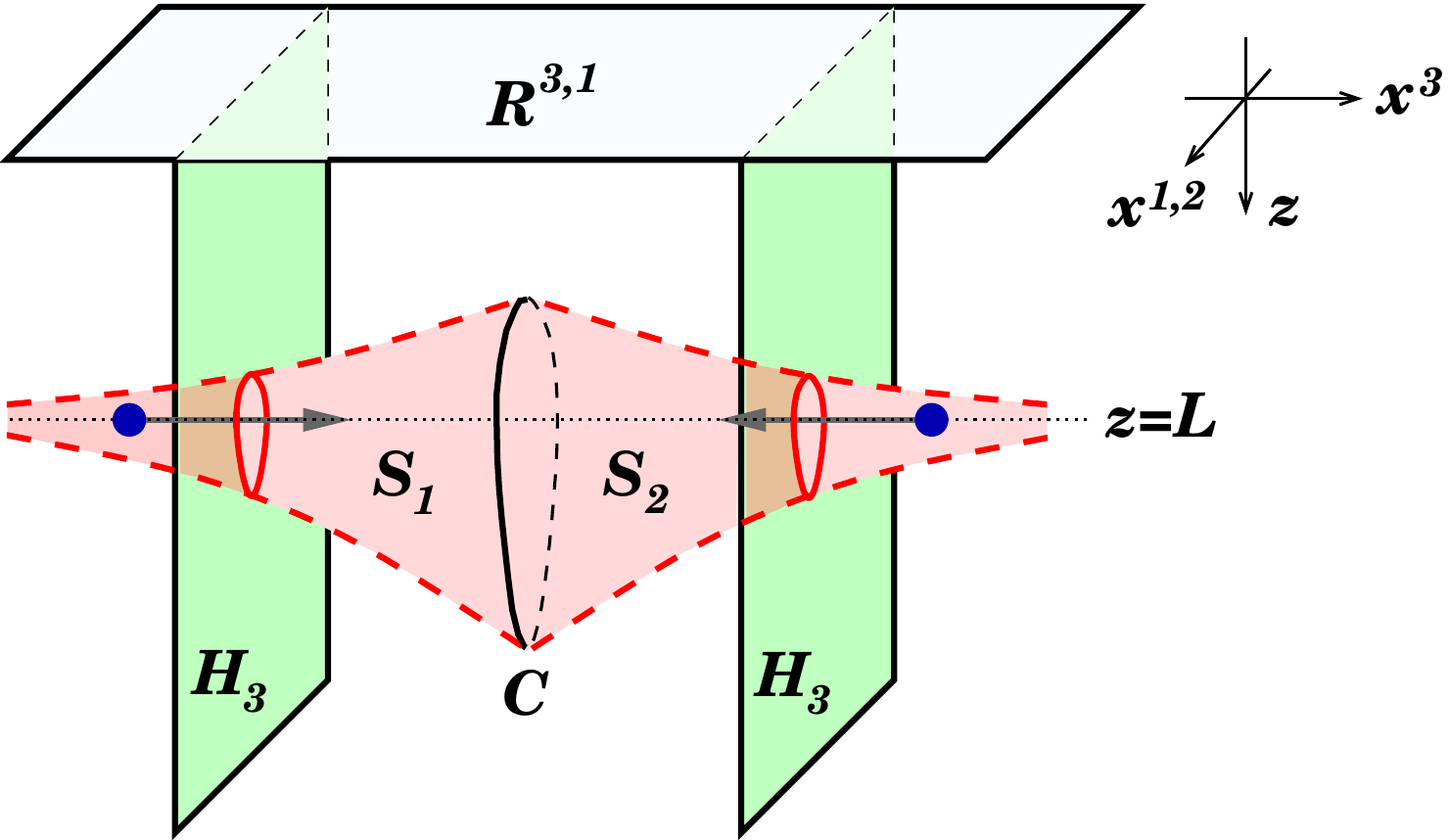}\hfill
   \includegraphics[width=2.5in]{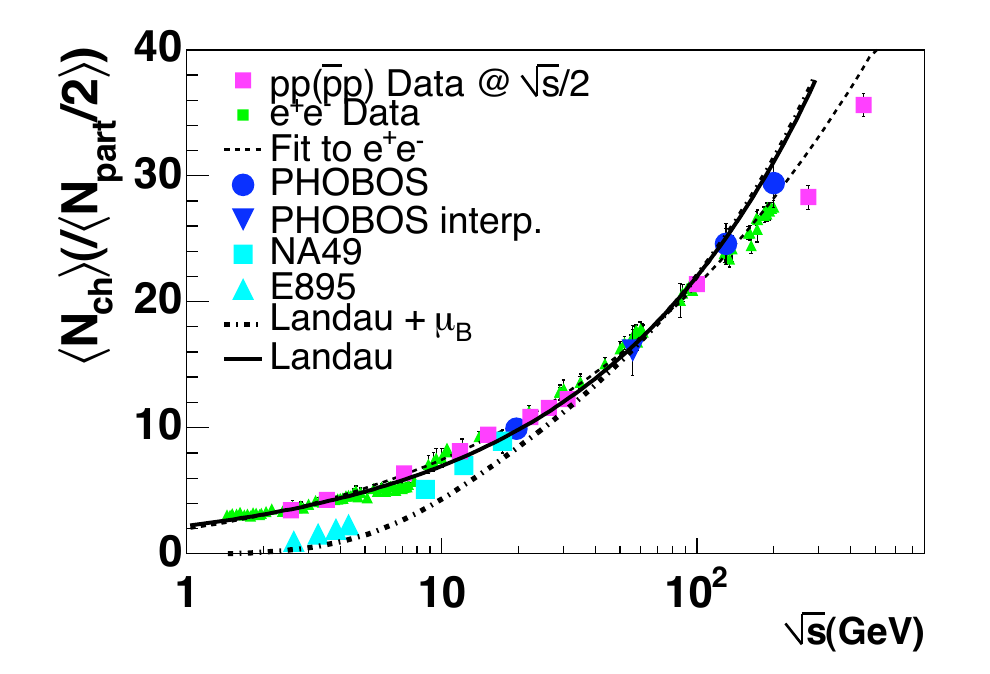}
  \caption[]{(Color online) Left: A trapped surface forms around the point-sources of two gravitational shock waves colliding head-on in $AdS_5$.  Right (from \cite{Steinberg:2004vy}): Total multiplicity scales approximately as $E^{1/2}$ (as predicted by the Landau model) over a wide range of energies.  It may just be starting to roll over to a slower scaling near top RHIC energies.}\label{TrapLandau}
 \end{figure}

The idea of colliding shocks is to replace a heavy ion with a boosted conformal soliton.  In the limit of infinite boost, the only non-zero component of the gauge theory stress tensor is
 \begin{equation}\label{BoostedSoliton}
  \langle T_{--} \rangle = {2EL \over \pi \left[
    (x^1)^2 + (x^2)^2 + L^2 \right]^3} \delta(x^-) \qquad\qquad
    L \approx 4.3\,{\rm fm} = \hbox{rms radius of gold} \,,
 \end{equation}
where $x^- = x^0-x^3$.  The fall-off of $\langle T_{--} \rangle$ at large $x_\perp$ is as a power, {\bf perilously} different from the exponential fall-off of the Woods-Saxon profile.  We nevertheless work with (\ref{BoostedSoliton}) because its gravitational dual is simple: it is a point-sourced gravitational shock wave in $AdS_5$.  As sketched in figure~\ref{TrapLandau}, a non-spherical event horizon forms when such shocks collide.
Following standard but non-rigorous methods, one can estimate
 \begin{equation}\label{Strapped}
  S \gsim \pi \left( {L^3 \over G_5} \right)^{1/3}
    (2EL)^{2/3} \approx 
    35000 \left( {\sqrt{s_{NN}} \over 200\,{\rm GeV}} \right)^{2/3}
     \,.
 \end{equation}
Using the phenomenological estimate $S \approx 7.5 N_{\rm charged}$, one gets $N_{\rm charged} \gsim 4700$ for central gold-gold collisions at top RHIC energies.  This compares quite favorably with the data, which gives $N_{\rm charged} \approx 5000$ \cite{Back:2004je}.  A {\bf serious peril} is the $E^{2/3}$ scaling in (\ref{Strapped}), which will bring the string theory prediction into conflict with data only slightly above top RHIC energies unless multiplicities rise unexpectedly quickly.

A crude resolution, which I nevertheless think is on the right track, is to discard the entropy of the part of the trapped surface above some ultraviolet cutoff in $AdS_5$.  This mimics the effect of asymptotic freedom and changes the scaling of $S_{\rm trapped}$ from $E^{2/3}$ to $E^{1/3}$, which is roughly in line with CGC predictions.  An infrared cutoff is probably also necessary.  With reasonable choices for the cutoffs, predictions for total multiplicity at LHC energies come out around $N_{\rm charged} = 20,000$.  This extrapolation is {\bf perilous} because it depends significantly on the cutoffs.

\section{Conclusions}\label{CONCLUSIONS}

The string theory approach to the quark-gluon plasma has made impressive progress.  AdS/CFT provides {\it many} calculations of strongly coupled phenomena that can be compared to heavy-ion collisions.  Such comparisons often come out surprisingly well, among them shear viscosity, the drag force on heavy quarks, jet splitting, total multiplicity at $\sqrt{s_{NN}} = 200\,{\rm GeV}$, and perhaps also thermalization and bulk viscosity.  (Due to lack of space I have been unable to include discussions of jet splitting and thermalization.)  When calculations in AdS/CFT compare poorly with QCD, we often understand why: Usually, it is the strong coupling limit and/or conformal invariance which distorts the results.

But the string theory approach is afflicted with significant {\bf perils}.  We are too often limited to the $N \gg 1$, $\lambda \gg 1$ regime.  ${\cal N}=4$ SYM isn't QCD, so comparisons aren't systematic.  We can go beyond ${\cal N}=4$ SYM in string theory, but the constructions become non-unique, providing room for theoretical fudging.  Essentially, this means that the onus is on theorists to create models that are as clean and predictive as possible while capturing the essentials of QCD.

Further effort on the theoretical side, and even better measurements of both hard probe and bulk physics observables, are clearly in order to clarify the extent to which the heavy-ion programs at RHIC and the LHC probe experimental predictions of string theory.

\section*{Acknowledgments}

I am particularly indebted to my collaborators: J.~Friess, G.~Michalogiorgakis, S.~Pufu, and A.~Yarom.  I thank J.~Casalderrey-Solana, M.~Gyulassy, B.~Jacak, J.~Nagle, J.~Noronha, K.~Rajagopal, D.~Teaney, and W.~Zajc for useful discussions.  This work was supported in part in part by the DOE under Grant No.\ DE-FG02-91ER40671 and by the NSF under award number PHY-0652782.

   


\def\href#1#2{#2}
\bibliographystyle{ssg}
\bibliography{pitp06}

\end{document}